\documentclass[lettersize,journal]{IEEEtran}
\pdfoutput=1
\usepackage{amsmath,amsfonts}
\usepackage[boxed]{algorithm2e}
\usepackage{array}
\usepackage{caption}
\usepackage{textcomp}
\usepackage{stfloats}
\usepackage{url}
\usepackage{verbatim}
\usepackage{graphicx}
\usepackage{cite}
\usepackage{braket}
\usepackage{qcircuit}
\usepackage{float}
\usepackage[table]{xcolor}
\usepackage{multirow, makecell,xcolor}

\begin{document}

\title{A Novel Fast Path Planning Approach for Mobile Devices using Hybrid Quantum Ant Colony Optimization Algorithm}

\author{
	\IEEEauthorblockN{%
	Mayukh Sarkar\IEEEauthorrefmark{1}, Jitesh Pradhan\IEEEauthorrefmark{1}, Anil Kumar Singh\IEEEauthorrefmark{2}, Hathiram Nenavath\IEEEauthorrefmark{3}} \\
	\IEEEauthorblockA{%
	\IEEEauthorrefmark{1}Department of Computer Science and Engineering, National Institute of Technology Jamshedpur, India \\
	\IEEEauthorrefmark{2}Department of Computer Science and Engineering, Motilal Nehru National Institute of Technology Allahabad, India\\
	\IEEEauthorrefmark{3}Department of Electronics and Communication Engineering, National Institute of Technology Jamshedpur, India }
}




\maketitle

\begin{abstract}
With IoT systems' increasing scale and complexity, maintenance of a large number of nodes using stationary devices is becoming increasingly difficult. Hence, mobile devices are being employed that can traverse through a set of target locations and provide the necessary services. In order to reduce energy consumption and time requirements, the devices are required to traverse following a Hamiltonian path. This problem can be formulated as a Travelling Salesman Problem (TSP), an NP-hard problem. Moreover, in emergency services, the devices must traverse in real-time, demanding speedy path planning from the TSP instance. Among the well-known optimization techniques for solving the TSP problem, Ant Colony Optimization has a good stronghold in providing good approximate solutions. Moreover, ACO not only provides near-optimal solutions for TSP instances but can also output optimal or near-optimal solutions for many other demanding hard optimization problems. However, to have a fast solution, the next node selection, which needs to consider all the neighbors for each selection, becomes a bottleneck in the path formation step. Moreover, classical computers are constrained to generate only \textit{pseudorandom} numbers. Both these problems can be solved using quantum computing techniques, i.e., the next node can be selected with proper randomization, respecting the provided set of probabilities in just a single execution and single measurement of a quantum circuit. Simulation results of the proposed Hybrid Quantum Ant Colony Optimization algorithm on several TSP instances have shown promising results, thus expecting the proposed work to be important in implementing real-time path planning in quantum-enabled mobile devices.
\end{abstract}


\section{Introduction}
\label{section-introduction}
As the scale and complexity of the IoT systems are increasing, the maintenance of a large number of stationary sensors at target locations is becoming costlier. Hence, mobile devices such as Unmanned Ground Vehicles and Unmanned Aerial Vehicles are being deployed. Rather than being stationary, these devices can traverse through target locations and provide the necessary services \cite{fu2022}. However, providing the services at target locations in real time requires extremely fast planning of possible paths. Moreover, the devices need to traverse as little distance as possible while visiting each target location. This problem can be formulated as a Travelling Salesman Problem (TSP), a well-known NP-hard problem.

The intractability of the Travelling Salesman Problem involves the applications of various optimization techniques that can provide near-optimal routes in a graph of target nodes. The most notable techniques among them include Ant Colony Optimization (ACO). Colorni et al. proposed the possibility of optimization following the natural phenomena occurring in ant colonies in 1991 \cite{colorni1991}. Dorigo and Gambardella successfully utilized this concept in optimizing the famous Travelling Salesman Problem in 1997 \cite{dorigo1997}. Since then, ACO has been utilized with great success in solving multitudes of problems that range from basic optimization up to as diverse as analog design \cite{okobiah2012}, reversible circuit synthesis \cite{sarkar2013}, etc.

Though ACO has been quite successful in obtaining near-optimum and optimum solutions for many optimization problems, including TSP, its execution needs to be sped up for real-time solutions such as path planning in mobile devices. One of the major bottlenecks in its speed, especially with a large number of nodes, is in selecting the next node. Any TSP-solving algorithm eventually needs to find the complete Hamiltonian cycle. However, while finding a path, selecting the next node among many possible neighbors needs to be executed fast to obtain the complete path as fast as possible. In the classical implementation of the ACO, finding the next neighbor involves calculating the cumulative probabilities of all neighbors and then selecting one based on a pseudorandom number. Quantum hardware can provide substantial speed-up in such scenarios by processing all probabilities together by a single execution of a quantum circuit, thanks to its inherent parallel nature obtained from superposition. Though the speed-up will be linear, with a large number of nodes and a sufficiently large number of ants, the overall accumulation of the speed-up starts to show its effects. This also has the advantage of selecting the next node based on pure randomization rather than using pseudorandomization as obtained from the classical counterpart. This paper attempts to achieve the same by proposing a novel hybrid classical-quantum technique of ACO.

Quantum computing started its journey from the proposal of a quantum equivalent of the Turing machine by Paul Benioff \cite{benioff1980}\cite{benioff1982}. In 1982, Richard Feynman postulated that simulating quantum physics on classical hardware would be impossible and suggested using quantum mechanical hardware for more efficient computation \cite{feynman1982}. The real power of quantum computing technique came into view when Peter Shor, in 1994, proposed quantum algorithms for prime factorization and discrete logarithm problems, hence showing the possibility of breaking RSA and Diffie-Hellman encryption techniques \cite{shor1994}. In 1996, Lov Grover showed the possibility of speed-up of search in an unstructured database using quantum system \cite{grover1996}. These results immediately brought quantum computing to the forefront of futuristic computational technique research. The declaration of quantum supremacy by Google using their Sycamore processor \cite{arute2019}, and the possibility of universal quantum logic operations with high average fidelity \cite{mkadzik2022} has raised the curtain of those possibilities which have been only theoretical concept for long. 

The current paper attempts to harness this power of quantum computation to speed up the well-known classical algorithm of ACO and employ it in possible real-time path planning applications as required in mobile devices. The rest of the paper is organized as follows. Section \ref{section-background} provides the necessary overview of quantum computing and the ant colony optimization algorithm. Several attempts have been made to provide a quantum version of ACO in literature. Section \ref{section-related-works} provides a brief overview of previous attempts and the novelties of the current paper. Section \ref{section-proposed-work} demonstrates the complete Hybrid Quantum Ant Colony Optimization algorithm as proposed in this paper, step-by-step, with an example of an instance of TSP. Section \ref{section-experiments} shows the performance of the proposed algorithm on several instances of the TSP obtained from a well-known collection of TSP instances, namely TSPLIB \cite{reinhelt2014}. The result shows that the proposed algorithm can provide optimal or near-optimal solutions to TSP problems with sufficient speed on real quantum hardware. Section \ref{section-conclusion} finally concludes the paper.

\section{Background Information}
\label{section-background}
In this section, required background information on quantum computing and ant colony optimization is being demonstrated.

\subsection{Quantum Computing}
\label{subsection-quantum-computing}
In classical computing, the basic unit of information, i.e., a bit, can stay in one of the two states, either 0 or 1, at any particular time. But in a quantum computer, a \textit{qubit} can stay as a \textit{superposition} of the two basis states $\ket{0}$ and $\ket{1}$, where state $\ket{0}$ corresponds to classical state 0 and state $\ket{1}$ corresponds to classical state 1. Mathematically, quantum state of a qubit can be represented as a vector $\ket{\psi} = \begin{bmatrix} \alpha \\ \beta \end{bmatrix} \in  \mathbb{C}^2 $, where $\alpha$ and $\beta$ are amplitudes associated with the states $\ket{0}$ and $\ket{1}$ respectively. Upon measurement, the quantum state collapses to one of the basis states and outputs the corresponding classical state. As for example, for the above-mentioned quantum state, upon measurement, the quantum state will collapse to state $\ket{0}$ and will output classical bit $0$ with probability $|\alpha|^2$, and it will collapse to state $\ket{1}$ and will output classical bit $1$ with probability $|\beta|^2$. Thus, the basis state $\ket{0}$ is associated with the vector $\ket{0} = \begin{bmatrix} 1 \\ 0 \end{bmatrix}$ and basis state $\ket{1}$ is associated with the vector $\ket{1} = \begin{bmatrix} 0 \\ 1 \end{bmatrix}$. An arbitrary quantum state $\ket{\psi}$ can thus be represented as follows.

\begin{equation*}
\ket{\psi} = \alpha \ket{0} + \beta \ket{1}
\end{equation*}

Moreover, as the measurement of a qubit should always result in either 0 or 1, it must follow that $|\alpha|^2 + |\beta|^2 = 1$.

On a multiqubit system, the qubits involved in the system, remain entangled to each other. As an example, on a two-qubit system, the state $\ket{\psi}$ becomes the superposition of all possible two-bit basis states $\ket{00}$, $\ket{01}$, $\ket{10}$ and $\ket{11}$, and can be written as follows.

\begin{equation*}
\ket{\psi} = \alpha_{00}\ket{00} + \alpha_{01}\ket{01} + \alpha_{10}\ket{10} + \alpha_{11}\ket{11}
\end{equation*}

, where $|\alpha_{00}|^2 + |\alpha_{01}|^2 + |\alpha_{10}|^2 + |\alpha_{11}|^2 = 1$. On measurement of both qubits, the classical result will be either $00$, $01$, $10$ or $11$ with the probabilities equal to the square of their corresponding amplitudes. Upon measuring a single qubit, as for example the first qubit, it will output $0$ with probability $|\alpha_{00}|^2 + |\alpha_{01}|^2$ and will output $1$ with probability $|\alpha_{10}|^2 + |\alpha_{11}|^2$. In case the first qubit measures to $0$, the complete quantum system collapses to the state $\frac{\alpha_{00}}{\sqrt{|\alpha_{00}|^2 + |\alpha_{01}|^2}} \ket{00} + \frac{\alpha_{01}}{\sqrt{|\alpha_{00}|^2 + |\alpha_{01}|^2}} \ket{01}$. The similar post-measurement state can be obtained in case the first qubit measures to $1$.

The state of a qubit gets modified by quantum gates, and a linear sequence of quantum gates eventually forms a quantum circuit. There cannot be any fanout or loop in a quantum circuit. The state of a quantum system can be manipulated using an adequately designed sequence of gates to harness the inherent parallelism obtained from the superposition property. A quantum gate must be a unitary matrix, and conversely, any unitary matrix is well-suited for a quantum gate. Only a small set of quantum gates is majorly used in quantum architectures, especially the universal set of quantum gates. There are uncountably many possible sets of universal quantum gates. As an example, a Clifford gate set $\lbrace H, S, CNOT \rbrace$, along with a non-Clifford gate such as $T$ gate, is a well-known universal gate set. Now, any single qubit gate $U$ can be represented as a sequence of three rotational gates up to a global phase factor, i.e., for any unitary $U$, there exist four real numbers $\alpha$, $\beta$, $\gamma$, and $\delta$, such that

\begin{equation*}
U = e^{i\alpha}R_z(\beta)R_y(\gamma)R_z(\delta)
\end{equation*}

As all gates in the mentioned gate set are single qubit gates except $CNOT$ gate, it can be said that any functionality can be implemented using two rotation gates and $CNOT$ gate. A complete overview of the basics of quantum computing, along with the proof of the above representation, can be found in \cite{nielsen2010}.

In the proposed work, a quantum circuit will be utilized to produce a quantum state provided the probabilities of nodes to be selected, and then will be measured to obtain the next node. As the probabilities involved in the ACO are real, the circuit will need to generate only those quantum states that are in real vector space. This allows us to restrict ourselves to only $R_y$ and $CNOT$ gates. The quantum selector proposed in \cite{sarkar2023} has been utilized here for the next node selection and is built up of only controlled-$R_y$ and $CNOT$ gates. An overview of the proposed quantum selector has been given in subsection \ref{subsection-quantum-selector}.

\subsection{Ant Colony Optimization}
\label{subsection-aco}
Ant Colony Optimization (ACO) is a well-known evolutionary algorithm that can obtain optimal or near-optimal solutions to many demanding optimization problems. This technique was first proposed in \cite{colorni1991}. In this technique, multiple ants move through some path in a graph in order to obtain optimal solutions. Each ant determines its path as follows.

\begin{enumerate}
\item If there can be multiple possible start nodes, the ant can start from any node randomly, or otherwise, it will start from the root. Then, it follows a path from the starting node as long as it does not reach some solution or as long as it does not reach the leaf node, depending on the problem statement. While sitting at a node, the ant always tries to find the optimum next node probabilistically. This probability of the subsequent nodes depends on two criteria: firstly, the likeliness of the path, which generally is some function of the path's cost. In a minimization problem, the likeliness varies inversely with the cost; in a maximization problem, the likeliness varies proportionately with the cost. The second selection criterion represents the experience of previous ants in finding paths for optimal or near-optimal solutions. For the first ant, this factor is initialized to the same value for all edges. This factor is commonly known as \textit{pheromone}. Now, let from the node $a$, the selection of the next node $b$ involves the likeliness $w_{ab}$ and the pheromone $\eta_{ab}$. Then, the probability of the node will be calculated as in equation \ref{eqtn-aco-next-node-prob}.

\begin{equation}
\label{eqtn-aco-next-node-prob}
P(a \rightarrow b) = \frac{w_{ab}^\alpha \eta_{ab}^\beta}{\sum_{i \in N(a)}w_{ai}^\alpha \eta_{ai}^\beta}
\end{equation}

, where $\alpha$ and $\beta$ are two hyperparameters that needs to be tuned for better results, and sum in the denominator involves all neighbours of node $a$ ($N(a)$ represents the neighbour set of node $a$).

\item Once the ant completes its tour, the edges' pheromones change along with the success of the ant following the rule as in equation \ref{eqtn-aco-pheromone-update}.

\begin{equation} \label{eqtn-aco-pheromone-update}
\eta_{ab} = (1 - \rho)*\eta_{ab} + f(w)
\end{equation}

, where $\rho$ is the evaporation rate of the paths, and $f(w)$ is the predetermined function of the overall cost of the final solution found if the ant is successful. If the ant fails to reach any solution, $f(w)$ is taken as zero. If the ant is successful, $f(w)$ varies inversely with $w$ for minimization problems and proportionately for maximization problems. In simpler terms, if the ant fails, the pheromones will only be evaporated, and nothing will be deposited. Furthermore, if the ant succeeds, some amount of pheromone gets deposited depending on the likeliness of the solution found, i.e., the better the solution, the higher the amount of pheromone deposited. For a large number of ants, this eventually allows all ants to converge toward the optimal solution by depositing more pheromones over the optimal paths.
\end{enumerate}

\section{Related Works}
\label{section-related-works}
Wang et al. \cite{wang2007}\cite{wang2008} introduced the novel concept of implementing Ant Colony Optimization on a Quantum Computer. In this work, the authors have implemented discrete binary ACO by representing the pheromone, i.e., the probability of the selected path, by qubits. A similar approach of quantum ACO has been employed in implementations of collision detection in virtual environment \cite{yang2011}, privacy-preserving data mining algorithm \cite{jue2013}, optimization of evacuation path \cite{zhang2013}, rule-based fuzzy classifier \cite{wu2015}. Wang et al., in the same paper \cite{wang2007}, have also proposed a quantum-inspired ACO employing the same approach, and has been utilized in link prediction \cite{cao2018}. A similar approach to quantum-inspired ACO has been employed in optimizing queries in distributed database systems \cite{mohsin2021}. The major problem of these works is the updation of the qubit by rotation gates, whose rotation angle is determined by a function of the probability amplitudes of the qubit being applied. This technique, thus, involves knowing the state of the qubit. However, on real quantum hardware, measuring a qubit would collapse its state, and hence, these algorithms cannot be implemented on a real quantum computer. 

A different approach to quantum implementation of ACO has been performed by representing the ant positions by coordinates on Bloch sphere \cite{chen2012}\cite{li2012}. A similar idea has also been employed in various optimization problems, such as in path planning of automated guided vehicles \cite{li2020}. However, the major problem with this approach is its lack of consistency with the original concept of Ant Colony Optimization.

Ghosh et al. \cite{ghosh2021} have proposed another novel approach to implementing ACO on a quantum computer named MNDAS. This paper employs a set of qubits to encode all paths, a separate set of qubits to represent pheromones, and three qubits as registers, requiring many qubits in a sufficiently large instance of graphs. Moreover, the proposed architecture in the paper employs a large number of SWAP gates, resulting in the tendency of introducing noise \cite{de2022}. Thus, such architecture is not implementable in current Noisy Intermediate-Scale Quantum (NISQ) era quantum computers.

De Andoin et al. \cite{de2022} have proposed an implementable ACO on a hybrid quantum system. The algorithm in this paper also does not follow the conventional approach of ACO and hence faces problems in constrained optimization problems. The paper proposes generating valid solutions from invalid ones in such cases, but that requires the knowledge of the complete set of valid solutions, which is impossible in most optimization problems.

In the current paper, the authors have proposed a novel implementation of ACO on a hybrid quantum computer that follows the exact steps of classical ACO. This gives the freedom of employing ACO in solving optimization problems, irrespective of being constrained or unconstrained, in the same manner as its classical counterparts. The proposed technique in this paper does not encode ants, pheromones, or paths using qubits like previous approaches in the literature. Instead, the arithmetic computations in the original ACO are assigned to a classical processor, and the quantum coprocessor makes the probabilistic selection of the next node. This design is novel because quantum computers are not known to provide sufficient speed improvements in arithmetic computations over classical ones. Moreover, in the present NISQ era, the fault-tolerance of classical computers is much more reliable than quantum computers. Along with possible speed-up in selecting the next node, employing a quantum computer in the subsequent node selection allows us to perform the selection with pure randomization rather than employing pseudorandom numbers as in classical computers.

\section{Proposed Implementation of Ant Colony Optimization on Hybrid Quantum System}
\label{section-proposed-work}
This section demonstrates the proposed implementation of the Ant Colony Optimization (ACO) algorithm on a hybrid Quantum-Classical system. Firstly, all the steps involved in the ACO algorithm can be implemented on a quantum computer because any classical algorithm can be simulated on a quantum computer efficiently using Fredkin Gate \cite{nielsen2010}. Moreover, several implementations of classical logical and arithmetic operations on a quantum computer have been proposed in literature \cite{ruiz2017}\cite{zhang2020}\cite{phillip2023}. However, in the current NISQ era, quantum computers are not fault-tolerant enough, and they can lose coherence before executing for sufficiently long \cite{brooks2019}. Thus, the algorithms employing a quantum-classical hybrid approach in solving problems rather than employing quantum hardware solely, such as Variational Quantum Eigensolver (VQE) \cite{peruzzo2014} and Quantum Approximation Optimization Algorithm (QAOA) \cite{farhi2014} have made immediate impact in this NISQ era of quantum computation.

The proposed algorithm in this paper employs the exact steps of ACO as its classical counterpart, except a suitable quantum circuit is performing the next node selection. Besides the possibility of implementing ACO on a futuristic NISQ respecting quantum system, the proposed Hybrid Quantum Ant Colony Optimization (HQACO) has the following advantages.

\begin{enumerate}
\item Possibility of solving optimization problems following the same approach as on classical computers. There is no need for some special preprocessing of input data other than the ones necessary for applying conventional ACO.

\item Employing a quantum computer in selecting the next node has two major advantages. In the classical counterpart, a \textit{pseudorandom} number is generated, and the cumulative probability of all possible next nodes is compared one by one with the generated number. The quantum selector will parallel process all possible next nodes in a single execution and select the next node with pure randomization respecting the exact probabilities. Thus, the quantum selector will speed up the next node selection and allow us to implement the ACO perfectly by enacting pure randomization rather than using pseudorandom numbers generated by classical random number generators.
\end{enumerate}

The current section elaborates on the proposed algorithm as follows. Firstly, subsection \ref{subsection-quantum-selector} demonstrates the design of a quantum selector that, upon providing a set of probabilities adding up to 1, will select one of the elements respecting their corresponding probabilities. The proposed HQACO algorithm utilizes this selector to select the next node. The complete step-by-step details of HQACO, with an example of the Travelling Salesman Problem, have been demonstrated in subsection \ref{subsection-hqaco-algo}.

\subsection{Design of the Quantum Selector}
\label{subsection-quantum-selector}

In \cite{sarkar2023}, the author has demonstrated an algorithm that given a set of probabilities $\lbrace p_i \rbrace$ such that $\sum p_i = 1$, will generate a parameterized quantum circuit generating the output state $\sum{\sqrt{p_i} \ket{i}}$. As an example, given set of probabilities $\lbrace 0.3, 0.2, 0.4, 0.1 \rbrace$, the algorithm will generate a two-qubit circuit producing the state $(\sqrt{0.3} \ket{00} + \sqrt{0.2} \ket{01} + \sqrt{0.4} \ket{10} + \sqrt{0.1} \ket{11})$. Observe that, upon measurement of the two qubits, the classical output \emph{00} will appear with probability \emph{0.3}, \emph{01} will appear with probability \emph{0.2}, \emph{10} will appear with probability \emph{0.4} and \emph{11} will appear with probability \emph{0.1}. In this subsection, a brief overview of the algorithm is being provided. Detailed discussion with its correctness can be found in \cite{sarkar2023}.

\begin{enumerate}
\item Generate spherical coordinates from the set of amplitudes employing algorithm \ref{algo-param-generator}. As for example, from the set of amplitudes $\lbrace \sqrt{0.3}, \sqrt{0.2}, \sqrt{0.4}, \sqrt{0.1} \rbrace$, the algorithm will generate the spherical coordinate angles $\lbrace r_1, r_2, r_3 \rbrace = \lbrace 1.982313, 2.013707, 0.927295 \rbrace$. It can be readily verified that

\begin{align*}
cos(\frac{r_1}{2}) &= \sqrt{0.3} \\
sin(\frac{r_1}{2})cos(\frac{r_2}{2}) &= \sqrt{0.2} \\
sin(\frac{r_1}{2})sin(\frac{r_2}{2})cos(\frac{r_3}{2}) &= \sqrt{0.4} \\
sin(\frac{r_1}{2})sin(\frac{r_2}{2})sin(\frac{r_3}{2}) &= \sqrt{0.1}
\end{align*}

\begin{algorithm}
\caption{GetParameters}
\label{algo-param-generator}
\SetKwInOut{Input}{input}
\SetKwInOut{Output}{output}
\Input{Array of amplitudes $A = \lbrace a_1, a_2, \cdots, a_n \rbrace$ of length $n$}
\Output{Array of spherical angles $R = \lbrace r_1, r_2, \cdots r_{n-1} \rbrace$ of length $n-1$}
\BlankLine
\For{$i\leftarrow 1$ \KwTo $n-1$}{
	\If{$a_i$ is $0$} {
		Append \emph{0} to $R$. \\
		Fill the rest of $R$ with $\pi$. \\
		break.
	}
	\Else {
		Set $r_i \leftarrow 2 * arccos(a_i)$. \\
		Divide all elements of $A$ by $sin(r_i / 2)$.
	}
}
\end{algorithm}

\item For single-qubit circuit generating 2-dimensional vector, the circuit will be, trivially a single $R_y$ gate with the angle set to the spherical angle generated in step 1. Observe that, from a 2-dimensional amplitude vector $\ket{\psi} = \begin{bmatrix} a_1 \\ a_2 \end{bmatrix}$, step 1 will generate a single angle $r_1$ such that $\begin{bmatrix} a_1 \\ a_2 \end{bmatrix} = \begin{bmatrix} cos(\frac{r_1}{2}) \\ sin(\frac{r_1}{2}) \end{bmatrix}$. The circuit is as shown in figure \ref{fig-single-qubit-generator}.

\begin{figure}[H]
\centering
\[
\begin{array}{c}
\Qcircuit @C=1em @R=1.6em {
		\lstick{\ket{0}} & \gate{R_y(r_1)} & \qw & \rstick{\ket{\psi}}
}
\end{array}
\]
\caption{Single-qubit arbitrary state generator}
\label{fig-single-qubit-generator}
\end{figure}

\item For two-qubit circuit generating 4-dimensional statevector $\ket{\psi} = [a_1\ a_2\ a_3\ a_4]^T$, first the three parameters $\lbrace r_1, r_2, r_3 \rbrace$ will be generated following algorithm \ref{algo-param-generator}. Then the circuit genrating $\ket{\psi}$ will be as in figure \ref{fig-two-qubit-generator}.

\begin{figure}[H]
\centering
\[
\begin{array}{c}
\Qcircuit @C=1em @R=1.6em {
		\lstick{} & \gate{R_y(r_1)} & \ctrl{1} & \gate{R_y(\pi+r_3)} & \qw \\
		\lstick{} & \qw & \gate{R_y(-r_2)} & \ctrl{-1} & \qw\gategroup{1}{5}{2}{5}{2.0em}{\}} & \rstick{\ket{\psi}}
		\inputgroupv{1}{2}{.8em}{.8em}{\ket{0}}
}
\end{array}
\]
\caption{Two-qubit arbitrary state generator}
\label{fig-two-qubit-generator}
\end{figure}
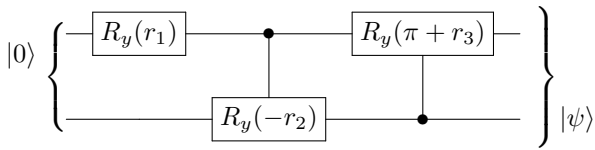

\item For any n-qubit circuit with $n > 2$ generating $2^n$-dimensional statevector $\ket{\psi}$, the $2^n-1$ dimensional vector of spherical angles will be generated following algorithm \ref{algo-param-generator}. The spherical angle vector will be provided as the $2^n-1$ parameters to the state generator circuit as shown in figure \ref{fig-n-qubit-generator}. In the circuit, $G_1$ is a recursive (n-1)-qubit arbitrary state generator circuit with first $(2^{n-1} - 1)$ parameters $\lbrace r_1, r_2, \cdots, r_{2^{n-1}-1} \rbrace$, the multicontrolled $R_y$ gate has all the above (n-1) lines as controls and employs $(2^{n-1})^{th}$ parameter. Next, there is a group of (n-1) CNOT gates, each having control on $n^{th}$ line and target to each of the above (n-1) lines respectively. Finally, there is another recursive (n-1)-qubit arbitrary state generator circuit $G_2$ with last $(2^{n-1} - 1)$ parameters $\lbrace r_{2^{n-1}+1}, \cdots, r_{2^{n}-1} \rbrace$. Each gate on $G_2$ has an additional control from the $n^{th}$ line.

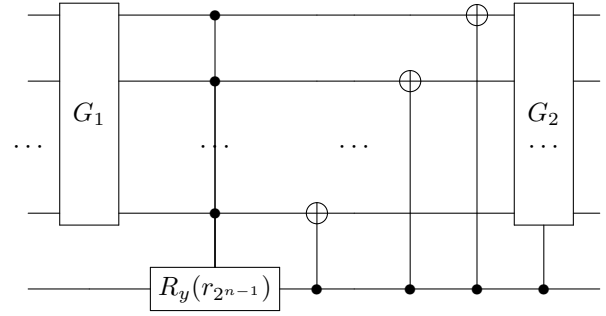
\begin{figure}[H]
\centering
\[
\begin{array}{c}
\Qcircuit @C=1em @R=1.6em {
		\lstick{} & \multigate{3}{G_1} & \ctrl{4} & \qw    & \qw & \qw          & \targ  & \multigate{3}{G_2} & \qw \\
		\lstick{} & \ghost{G_1}        & \ctrl{3} & \qw    & \qw & \targ        & \qw    & \ghost{G_2}        & \qw \\
		\cdots    & \nghost{G(_1}      & \cdots   & & \cdots                & \nghost{G_2}   &    & \cdots & \\
		\lstick{} & \ghost{G_1}        & \ctrl{1} & \targ  & \qw  & \qw        & \qw    & \ghost{G_2}        & \qw \\
		\lstick{} & \qw                & \gate{R_y(r_{2^{n-1}})} & \ctrl{-1} & \qw & \ctrl{-3} & \ctrl{-4} & \ctrl{-1} & \qw 
}
\end{array}
\]
\caption{Multiqubit arbitrary state generator}
\label{fig-n-qubit-generator}
\end{figure}

\end{enumerate}

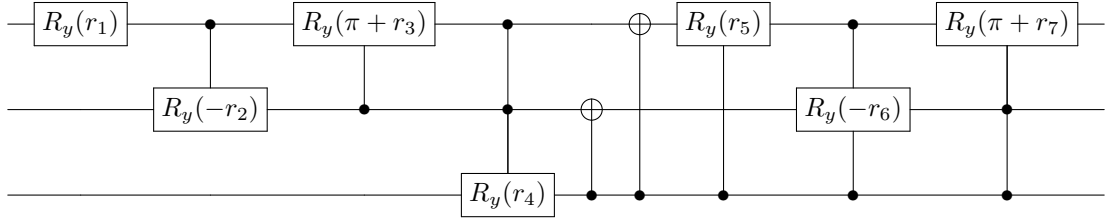
\begin{figure*}[h]
\centering
\[
\begin{array}{c}
\Qcircuit @C=1em @R=1.6em {
	\lstick{} & \gate{R_y(r_1)} & \ctrl{1} & \gate{R_y(\pi+r_3)}  & \ctrl{2} & \qw & \targ & \gate{R_y(r_5)} & \ctrl{1} & \gate{R_y(\pi+r_7)} & \qw \\
		\lstick{} & \qw & \gate{R_y(-r_2)} & \ctrl{-1} & \ctrl{1} & \targ & \qw & \qw & \gate{R_y(-r_6)}& \ctrl{-1} & \qw \\
		\lstick{} & \qw & \qw & \qw & \gate{R_y(r_4)} & \ctrl{-1} & \ctrl{-2} & \ctrl{-2} & \ctrl{-1} & \ctrl{-2} & \qw
}
\end{array}
\]
\caption{Example of 3-qubit arbitrary state generator}
\label{fig-3-qubit-generator}
\end{figure*}

Employing the stated algorithm, a 3-qubit arbitrary state generator circuit generating statevector $\lbrace a_1, a_2, \cdots, a_8 \rbrace$ can be designed as in figure \ref{fig-3-qubit-generator}, where the parameters generated from the input statevector, employing algorithm \ref{algo-param-generator} is $\lbrace r_1, r_2, \cdots, r_7 \rbrace$. The circuit generated using the stated algorithm can generate statevectors with only real amplitudes, which will be sufficient for the proposed HQACO algorithm, as the probabilites generated in the intermediate stages are always real numbers.

\subsection{HQACO Algorithm}
\label{subsection-hqaco-algo}

\begin{figure}[H]
\centering
\includegraphics[width=0.4\textwidth]{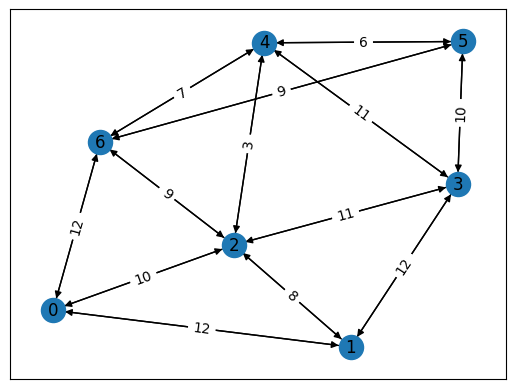}
\caption{An Example Graph for HQACO}
\label{fig-example-graph}
\end{figure}

In this subsection, the quantum selector circuit designed in subsection \ref{subsection-quantum-selector} will be employed to perform Ant Colony Optimization on a hybrid quantum computer. In demonstrating the algorithm, an instance of the Travelling Salesman Problem on a small graph, as shown in figure \ref{fig-example-graph}, will be employed. 

\begin{enumerate}
\item A parametrized quantum selector with $q = max(\lceil log_2(v) \rceil, \lceil log_2(d) \rceil)$ qubits needs to be prepared with the structure as mentioned in subsection \ref{subsection-quantum-selector}, where $v$ is the number of nodes, and $d$ is the maximum degree (outdegree for directed graph) of the graph. The given example graph has seven vertices, and the maximum degree is 4. Thus, it is sufficient to have a 3-qubit quantum selector, as depicted in figure \ref{fig-3-qubit-generator}. If the number of possible candidates, and hence, the number of probabilities to choose from, is less than $2^q$ (8, in the example), append 0 at the end of the amplitude array to make it of length $2^q$.

\item The graph information is stored in a classical register $C_1$. Each edge has to be initialized with a pheromone value, which will be kept in another classical register $C_2$. Initially, all edges receive the same pheromone value. The initial pheromone value has been taken as $0.5$ for all the edges in the given example.

\item Suitable values of the hyperparameters $\alpha$, $\beta$, and pheromone evaporation rate $\rho$ are chosen and kept in classical register $C_3$. In the given example, let us choose the values $\alpha=0.4$, $\beta=0.6$ and $\rho=0.01$ respectively. This will allow the selection to be more directed by pheromone values rather than edge costs.

\item Perform the following sequence of operations for each ant.

\begin{enumerate}
\item The start node for the ant is to be selected with equal probability. Set the parameters of the quantum selector so that the probability of each node becomes $\frac{1}{v}$. In the example with $7$ nodes, the $7$ parameters of the quantum selector will be the angles obtained using algorithm \ref{algo-param-generator} from the set of amplitudes $\lbrace \sqrt{\frac{1}{7}}, \sqrt{\frac{1}{7}}, \sqrt{\frac{1}{7}}, \sqrt{\frac{1}{7}}, \sqrt{\frac{1}{7}}, \sqrt{\frac{1}{7}}, \sqrt{\frac{1}{7}}, 0 \rbrace$. After calculating the parameters, the classical processor will set them to the quantum selector circuit for further processing.

\item Quantum registers are initialized with $\ket{0}$. The quantum selector will produce the desired state. All the qubits of the selector will be measured and stored in a classical register $C_3$. The classical register holds the start node index with equal probabilities. Suppose the initial node has been measured as $4$ in the example considered.

\item \label{step-1} The classical processor will read the node from the register $C_3$ and calculate the probabilities of the next transitions. To calculate this, the classical processor will store all unvisited neighbors and consider the inverse of the corresponding edge costs as the corresponding weights. The probability of a particular incident edge $e$ is then evaluated as $\frac{w(e)^\alpha ph(e)^\beta}{\sum_{i=1}^d w(e)^\alpha ph(e)^\beta}$, where $w(e)$ and $ph(e)$ are the edge weights and current pheromone values respectively, and $d$ is the degree of the current node. As for example, with start node $4$ and all pheromone values $0.5$, the probability of going to neighbor nodes $\lbrace 2, 3, 5, 6 \rbrace$ are $\lbrace 0.32625, 0.19402, 0.24725, 0.23247 \rbrace$, as the corresponding edge costs are $\lbrace \frac{1}{3}, \frac{1}{11}, \frac{1}{6}, \frac{1}{7} \rbrace$ respectively. Observe that the cost has been taken as the inverse of the edge weight, as it is a minimization problem.

\item \label{step-2} Evaluate the angles using algorithm \ref{algo-param-generator} from the zero-appended amplitude set, where each amplitude is the square root of the corresponding probabilities. Thus, in the continuing example, the classical processor will set the parameters of the quantum selector after evaluating them from the amplitude set $\lbrace 0.57118, 0.44048, 0.49724, 0.48215, 0, 0, 0, 0 \rbrace$.

\item \label{step-3} Reset the qubits in the quantum selector to $\ket{0}$, and allow them to pass through the quantum selector with set parameters. At the end, the qubits will be measured to obtain the next node index. For example, if the measurement result turns out to be $010$, the node at index 2 (0-indexed), i.e., node $5$, will be selected as the next node. Set the index in classical register $C_3$.

\item Continue the steps \ref{step-1} to \ref{step-3}, as long as no unvisited next node can be found. After completion of the tour, reduce the pheromone of each edge by pheromone evaporation rate $\rho$.

\item Two events can occur at this point: firstly, the ant has visited all nodes, in which case the ant is said to be successful, or secondly, the ant is stuck without any unvisited neighbor, but all nodes have not been visited. The tour is a failure in the second scenario, and the result will be discarded. In the first scenario, increase each edge's pheromone on the successful path by $\frac{1}{c}$, where $c$ is the total cost of the path.

\item If a Hamiltonian cycle is to be obtained, another condition needs to be added for a path to be successful, and that is, the last node on the path must have the starting node as its neighbor.
\end{enumerate}

In the given example, the above step has been performed with 100 ants, and the minimum Hamiltonian cycle $\lbrace 6, 5, 3, 1, 0, 2, 4, 6 \rbrace$ has been obtained with cost of 63.

\end{enumerate}

\subsection{Expansion to parallel ant systems}
\label{subsection-expansion-parallel}
The proposed HQACO algorithm can be expanded naturally to parallel ant systems, where multiple ants will start in parallel rather than sequentially at each iteration. On a classical computer, this is achieved by multithreads, which will be executed on multiprocessors or multicores, whichever is available. On a quantum computer, the same can be achieved by placing multiple quantum co-processors, each co-processor being selected by each ant in parallel. Thus, in such a system, each ant will have a separate unit of classical-quantum hybrid processor.

\section{Experimental Results}
\label{section-experiments}
This section demonstrates the experimental results of the proposed HQACO algorithm on several instances of the Travelling Salesman Problem obtained from TSPLIB \cite{reinhelt2014}, a well-known collection of TSP instances for experimentations. The experiment has been performed using Python 3.11 with Qiskit 0.43 in a locally installed Linux environment. For larger instances, the Qiskit provider for Amazon Braket has been used and implemented on Amazon AWS. The table \ref{table-experiment} demonstrates the experimental results for 5 TSP instances, with the minimum path found and the overall estimation error. In each instance, 1000 ants have been used with $\alpha=0.4$, $\beta=0.6$ and $\rho=0.01$. Ten shots have been used during each measurement, and the output with the highest appearance count has been selected, as a large number of shots would make the output nearly deterministic by driving the output towards the highest probability.

\begin{table*}[htbp]
\centering
\begin{tabular}{||c|c|c|c|c|c|c||}
\hline 
\multirow{2}{*}{\textbf{\thead{TSP\\Instance}}}
& \multirow{2}{*}{\textbf{\thead{Number\\of Nodes}}}
& \multirow{2}{*}{\textbf{\thead{Lower Bound\\In Specification \cite{reinhelt2014}}}}
& \multicolumn{2}{c|}{\textbf{\thead{Minimum Cycle Found using HQACO}}}
& \multirow{2}{*}{\textbf{\thead{Error\\in Estimation}}}
& \multirow{2}{*}{\textbf{\thead{Number of\\Qubits used}}}\\
\cline{4-5}
&&& \textbf{\thead{Cost}} & \textbf{\thead{Path}} && \\
\hline
\multirow{2}{*}{gr17.tsp} & \multirow{2}{*}{17} & \multirow{2}{*}{2085} & \multirow{2}{*}{2130} & [12, 16, 6, 7, 5, 13, 14, 2, 10,  & \multirow{2}{*}{2.16\%} & 5 for node selection \\
&&&& 9, 1, 4, 8, 11, 15, 3, 0, 12] && 4 for path selection \\
\hline
\multirow{2}{*}{burma14.tsp} & \multirow{2}{*}{14} & \multirow{2}{*}{3323} & \multirow{2}{*}{3547} & [12, 6, 11, 5, 4, 3, 2, 13,  & \multirow{2}{*}{6.74\%} & \multirow{2}{*}{4} \\
&&&& 1, 9, 10, 8, 7, 0, 12] && \\
\hline
\multirow{2}{*}{gr21.tsp} & \multirow{2}{*}{21} & \multirow{2}{*}{2707} & \multirow{2}{*}{3216} & [2, 8, 4, 15, 11, 3, 0, 5, 7, 6, 18,  & \multirow{2}{*}{18.8\%} &\multirow{2}{*}{5} \\
&&&& 19, 10, 16, 9, 17, 12, 13, 14, 1, 20, 2] && \\
\hline
\multirow{2}{*}{bayg29.tsp} & \multirow{2}{*}{29} & \multirow{2}{*}{1610} & \multirow{2}{*}{1922} & [23, 12, 15, 6, 22, 7, 27, 0, 5, 11, 8, 2, 28, 25, 4,  & \multirow{2}{*}{19.38\%} & \multirow{2}{*}{5} \\
&&&& 20, 1, 19, 9, 3, 14, 17, 13, 21, 16, 18, 24, 10, 26, 23] && \\
\hline
\multirow{2}{*}{bays29.tsp} & \multirow{2}{*}{29} & \multirow{2}{*}{2020} & \multirow{2}{*}{2599} & [6, 24, 18, 15, 14, 10, 13, 21, 16, 17, 3, 9, 19, 4,  & \multirow{2}{*}{28.66\%} & \multirow{2}{*}{5} \\
&&&& 8, 11, 5, 27, 0, 23, 7, 22, 25, 28, 2, 1, 20, 12, 26, 6] && \\

\hline
\end{tabular}
\caption{Experimental Results}
\label{table-experiment}
\end{table*}

In the first instance with 17 nodes, for each ant, a 5-qubit circuit is required to find the starting node, whereas a 4-qubit circuit is sufficient to find the path as each node has a degree of 16. In the simulation experiment, two separate circuits have been used for the same. The same can be followed and suggested in real quantum hardware rather than using a single 5-qubit circuit. Though it will increase the build cost, using a 5-qubit circuit, even where a 4-qubit circuit is sufficient, will increase the execution time and probability of noise in output.

The results have been sorted based on errors in estimation, and it can be observed that the error percentage increases with the number of qubits in the circuit. This output is reasonably expected, as the larger instances require more ants to converge, and the first 1000 ants in each case have shown apparent convergence towards a lower value. In the simulation, though larger circuits require exponential time to execute, they are expected to run extremely fast on real quantum hardware, as quantum simulation on classical hardware is exponential.
Hence, on real quantum hardware, a much larger number of ants can be utilized to obtain better results.

The above discussion shows that on real quantum hardware, the proposed HQACO algorithm can obtain near-optimum results of TSP instances quite fast, which is necessary for real-time path planning of mobile agents.

\section{Conclusions}
\label{section-conclusion}
This paper proposes a novel implementation of the Ant Colony Optimization algorithm on a hybrid quantum computer. This algorithm alleviates two essential problems in implementing the same on a classical computer. Firstly, during path formation, each ant selects the next node probabilistically, which involves calculations of the cumulative probability of each node, making it linear in time. Secondly, the random number generated during node selection is not a pure random number, as classical computers can generate only \textit{pseudorandom} numbers. Current work performs all arithmetic calculations, such as probability calculations, on a classical computer. Quantum computer in the NISQ era is not well-known enough to perform speedier arithmetic calculation reliably. Thus, the quantum phenomena have been applied at exactly that step where it would shine, i.e., the selection of the next node. A single execution of the quantum selector circuit, followed by measurement, will select the next node, respecting the assigned probabilities with pure randomness. Experimental results have shown that the speedy quantum circuit can reliably solve TSP instances with a suitable number of ants. Thus, this work is expected to be important in those areas where fast and accurate execution of ACO, such as solving TSP instances in real time for path planning of mobile devices.

\section{Acknowledgements}
\label{section-acknowledgements}
This work would not have been possible without the financial support of MeitY Quantum Computing Applications Lab (QCAL). The authors are also thankful to Amazon AWS for providing the excellent environment for Quantum Computing research needed to perform the experiments using the Qiskit provider for Amazon Braket.

\bibliography{references.bib}
\bibliographystyle{IEEEtran}

\vfill

\end{document}